\documentclass[10pt,a4paper]{article}

\usepackage{amsmath,amssymb}
\usepackage[dvipdfmx]{graphicx,psfrag}
\usepackage{latexsym}
\usepackage{slashed}

\def\lagrangian{\mathcal{L}}
\def\pd{\partial}

\makeatletter
\@addtoreset{equation}{section}
\makeatother

\usepackage[top=30truemm,bottom=30truemm,left=25truemm,right=25truemm]{geometry}

\title{Gauged Nambu-Jona-Lasinio inflation}
\author{Tomohiro Inagaki${}^{1,2}$, Sergei~D.~Odintsov${}^{3,4,5}$, and Hiroki~Sakamoto${}^5$,\\[2mm]
\normalsize
${}^1$Information Media Center, Hiroshima University, Higashi-Hiroshima, 739-8521, Japan,\\\normalsize
${}^2$Core of Research for the Energetic Universe, Hiroshima University, Higashi-Hiroshima,\\\normalsize
 739-8526,  Japan,\\\normalsize
${}^3$Instituci\'{o} Catalana de Recerca i Estudis Avan\c{c}ats (ICREA), Barcelona, Spain,\\\normalsize
${}^4$Consejo Superior de Investigaciones Cient\'{i}ficas, ICE/CSIC-IEEC, Campus UAB, \\\normalsize
Torre C5-Parell-2a pl, E-08193 Bellaterra (Barcelona), Spain,\\\normalsize
${}^5$Lab. Theor. Cosmology, Tomsk State University of Control Systems and Radioelectronics\\\normalsize
 (TUSUR), 634050 Tomsk, Russia,\\\normalsize
${}^5$Department of Physics, Hiroshima University, Higashi-Hiroshima, 739-8526, Japan\\}
\begin{document}
\maketitle

 \begin{abstract}
We investigate the gauged Nambu-Jona-Lasinio model  in curved spacetime at the large $N_c$ limit and in slow-roll approximation.
The model can be described by the renormalization group corrected gauge-Higgs-Yukawa theory with the corresponding compositeness conditions.
Evaluating the renormalization group (RG) improved effective action, we show that such model can produce CMB fluctuations and find inflationary parameters: spectral index, tensor-to-scalar-ratio and running of the spectral index.
 We demonstrate that the model can naturally satisfy the Planck 2015 data and maybe considered as an alternative candidate for Higgs inflation.\\
{\it Keyword:} gauged Nambu-Jona-Lasinio model, CMB fluctuations\\
{\it PACS:} 04.62.+v 12.60.Rc 98.80.Cq
 \end{abstract}

\section{Introduction}
The standard model of the particle physics is  based on the gauge symmetry and the symmetry breaking. For example, QCD is described as $SU(3)$ gauge theory. The gauge interaction causes chiral symmetry breaking. Because of the strong coupling constant, low energy phenomena are often approximated by using effective models. For instance, Nambu-Jona-Lasinio (NJL) model \cite{Nambu:1961} is a well-known model for light mesons.
Gauged NJL model maybe applied to the description of QCD, as $SU(N_c)$ gauge theory, in which the chiral symmetry is broken and meson like composite fields are generated at high energy scale (see, for example \cite{Miransky:1993}). The renormalizability of the model is precisely studied by using the ladder Schwinger-Dyson equation in flat spacetime \cite{Kondo:1991}. In the leading order of the $1/N_c$ expansion the model can be described by a gauge-Higgs-Yukawa type Lagrangian with compositeness conditions \cite{Leung:1985}.

Gauged NJL model is expected to be very important at the early universe\footnote{Constraints from the high energy collider physics are reviewed in Ref.~\cite{Hill:2002ap}.}. The vacuum energy for a fermion field is negative, unlike to the scalar field. After the dynamical symmetry breaking a composite scalar  can develop a positive vacuum energy. A primordial potential energy for the composite scalar  may induce inflation, i.e. exponential expansion of our universe. An evidence of inflation is observed in the cosmic microwave background radiation (CMB) \cite{Ade:2015tva,Ade:2015lrj}. In this paper we propose gauged NJL inflation model as an alternative to Higgs inflation. We  evaluate the CMB fluctuations in the gauged NJL model and demonstrate that inflationary parameters  are consistent with Planck bounds.

The study of NJL model in curved space-time has been started some time ago (for review, see Ref.~\cite{Inagaki:1993kz}). The curvature coupling for the composite scalar field has been obtained in Ref.~\cite{Hill:1991jc,Muta:1991mw}. In the gauged NJL model it has been shown that the chiral symmetry can be broken for a strong curvature \cite{Geyer:1996wg}. In Ref.~\cite{Inagaki:2011jk} the authors have discussed the contributions from the dynamical symmetry breaking to the expansion of the universe. However, as far as we know there is only preliminary work to argue a contribution from composite fields to the CMB fluctuations. For instance, in Ref.~\cite{Iso:2014gka} the authors have assumed non-vanishing fermion condensation and evaluated the CMB fluctuations.

This article is organized as follows: In Sec.~\ref{gNJL} the gauged NJL model is introduced. The model is rewritten as renormalization-group corrected gauge-Higgs-Yukawa Lagrangian with compositeness conditions.
In this way, this looks qualitatively similar to the study of RG improved effective potential in curved spacetime \cite{eli} and corresponding RG improved inflation \cite{rgi}. In Sec.~\ref{InflationaryParameters} we calculate the inflationary parameters in the gauged NJL model under the slow -roll approximation scenario. In Sec.~\ref{NumericalResults} the inflationary parameters are numerically calculated and compared with the Planck data. Then we find a possible parameters range consistent with Planck data for the gauged NJL model. In Sec.~\ref{EndInflation} we briefly discuss an exit from inflation. Some concluding remarks are given in Sec.~\ref{Conclusion}.

\section{Gauged NJL model\label{gNJL}}

We consider $N_f$ fermion flavors   with $SU(N_c)\otimes {\cal G}$ gauge symmetry. It is assumed that the gauge interaction for $\cal{G}$ is strong enough and well described by four-fermion interaction. We employ the gauged NJL model as an effective model of the $SU(N_c)\otimes {\cal G}$ gauge theory. The Lagrangian is given by
\begin{align}
\lagrangian_{gNJL} = \lagrangian_{gauge} + \bar{\psi} i\hat{\slashed{D}}\psi + \frac{16\pi^2 g_4}{8 N_f N_c \Lambda^2} \left[\left(\bar{\psi}\psi\right)^2+\left(\bar{\psi}i\gamma_5 \tau^{a}\psi\right)^2\right] ,
\label{L:gNJL}
\end{align}
where $\tau^a$ are the generators for the flavor symmetry, $\lagrangian_{gauge}$ shows the Lagrangian of the $SU(N_c)$ pure gauge theory and $\hat{D}$ is the covariant derivative for the $SU(N_c)$ gauge invariant and generally covariant fermion kinetic term. The gauge field for the gauge symmetry ${\cal G}$ is integrated out. We introduce the compositeness scale $\Lambda$ and use it to define the dimensionless four-fermion coupling $g_4$. It is noted that the Lagrangian has the $U(N_f)_L\otimes U(N_f)_R$ flavor symmetry.
Thus the model is regarded as a scale up of QCD with QED interaction. The fermion fields, $\psi$, are confined in baryon like states above the compositeness scale. We consider the model whose compositeness scale is higher than the one for inflation. The composite fields dynamically acquire compositeness scale mass. Only the possible environment to test the model is found at early universe.

According to the auxiliary field method the Lagrangian ($\ref{L:gNJL}$) is rewritten as
\begin{align}
\lagrangian_{aux} = \lagrangian_{gauge} + \bar{\psi} \left(i\hat{\slashed{D}}-\sigma-i\gamma_5 \tau^a\pi^a \right)\psi - \frac{2 N_f N_c\Lambda^2}{16\pi^2 g_4} \left(\sigma^2+{\pi^a}^2\right) .
\label{L:auxi}
\end{align}
Replacing the auxiliary fields $\sigma$ and $\pi^a$ by the solutions of the equations of motion,
\begin{align}
\sigma=-\frac{16\pi^2 g_4}{ 4 N_f N_c \Lambda^2} \bar{\psi}\psi,\,\,\ \pi^a=-\frac{16\pi^2 g_4}{ 4 N_f N_c \Lambda^2} \bar{\psi}i\gamma_5 \tau^{a}\psi,
\end{align}
 the original Lagrangian ($\ref{L:gNJL}$) is reproduced. Starting from the Lagrangian ($\ref{L:gNJL}$) at the compositeness scale $\Lambda$, one obtains the effective Lagrangian with a gauge-Higgs-Yukawa form below the compositeness scale through the renormalization group evolution \cite{Leung:1985},
\begin{align}
        \lagrangian_{gHY} = -\frac{1}{4}F^{\mu\nu}F_{\mu\nu} + \frac{1}{2}\pd_\mu\sigma\pd^\mu\sigma
        +\frac{1}{2}\pd_\mu\pi^a\pd^\mu\pi^a
        - \frac{1}{2}m^2(\sigma^2+\pi^a\pi^a) - \frac{\lambda}{4}(\sigma^2+\pi^a\pi^a)^2 \nonumber \\
        -\frac{1}{2}\xi R(\sigma^2+\pi^a\pi^a) + \bar{\psi} i\hat{\slashed{D}}\psi - y \bar{\psi}(\sigma+i\gamma_5 \tau^a\pi^a )\psi.
\label{L:gHY}
\end{align}
Because of the $U(N_f)_L\otimes U(N_f)_R$ flavor symmetry one can assume that only the composite scalar field, $\sigma$, contributes the vacuum energy and induce exponential expansion of the universe. Below we drop the pseudo-scalar field, $\pi^a$.

The RG equations for the gauge coupling $g(t)$, the Yukawa coupling $y(t)$, the quartic scalar coupling $\lambda(t)$ and the curvature coupling $\xi(t)$ are found  in the leading order with respect to a modified $1/N_c$ expansion \cite{Geyer:1996wg,Harada:1994wy},
\begin{align}
        &16\pi^2 \frac{d}{dt} g(t) = -bg^3(t),\nonumber \\
        &16\pi^2 \frac{d}{dt} y(t) = y(t)[ay^2(t)-cg^2(t)],\nonumber \\
        &16\pi^2 \frac{d}{dt} \lambda(t) = 4a y^2(t)[\lambda(t)-y^2(t)],\nonumber\\
        &16\pi^2 \frac{d}{dt} \xi(t) = 2ay^2(t)\left[\xi(t) - \frac{1}{6}\right],
\label{eq:RG}
\end{align}
where RG parameter $t$ is defined by $t\equiv\ln(\mu/\mu_0)$ with the renormalization scale $\mu$ and a reference scale $\mu_0$, $a,b,c$ and $u$ being the positive constants,
\begin{align}
a\equiv 2N_fN_c, \, b\equiv\frac{1}{3}(11N_c-2N_f), \, c\equiv 3\frac{N_c^2-1}{N_c}.
\end{align}
In Ref.~\cite{Bardeen:1989ds} it has been shown that the Lagrangian (\ref{L:gHY}) is identified with (\ref{L:auxi}) under the compositeness conditions in a flat space-time,
\begin{align}
        \frac{1}{y^2(t_\Lambda)} = 0,\,
        \frac{\lambda(t_\Lambda)}{y^4(t_\Lambda)} = 0,\,
        \frac{m^2(t_\Lambda)}{y^2(t_\Lambda)} = \frac{2a}{16\pi^2}\Lambda^2\left(\frac{1}{g_4}-\frac{1}{\Omega(t_\Lambda)}\right),
\label{cond:Composite}
\end{align}
where $\Omega(t_\Lambda)$ is a function of $t$ and it has a constant value $\omega$ at the limit, $b\to +0$ \cite{Harada:1994wy}.
It is shown that the composite scalar is conformally coupled to curvature \cite{Hill:1991jc}.
Thus the compositeness condition for $\xi$ is given as,
\begin{align}
        \xi(t_\Lambda) = \frac{1}{6}.\,
\label{cond:Composite:xi}
\end{align}

Solving the RG equations (\ref{eq:RG}) with the compositeness conditions (\ref{cond:Composite}) and (\ref{cond:Composite:xi}), we obtain the running couplings,
\begin{align}
        &y^2(t) = \frac{c - b}{a}g^2(t)\left[1 - \left(\frac{\alpha(t)}{\alpha(t_\Lambda)}\right)^{1 - c/b}\right]^{-1} \equiv y_\Lambda^2(t),
\label{running:y}\\
        &\frac{\lambda(t)}{y^4(t)} = \frac{2a}{2c - b}\frac{1}{g^4(t)}\left[1 - \left(\frac{\alpha(t)}{\alpha(t_\Lambda)}\right)^{1 - 2c/b}\right]
        \equiv \frac{\lambda_\Lambda(t)}{y_\Lambda^4(t)},
\label{running:lambda}\\
        &\xi(t) = \frac{1}{6} ,
\label{running:xi}
\end{align}
where $\alpha(t)$ is the running fine structure constant for the $SU(N_c)$ gauge interaction, $\alpha(t)\equiv g^2(t)/(4\pi)$.

 At the fixed gauge coupling limit, $b\to +0$, Eqs.~(\ref{running:y}) and (\ref{running:lambda}) reduce to
 \begin{align}
        &y_\Lambda^2(t) = \frac{16\pi^2}{2a}\frac{\alpha}{\alpha_c}\left[1 - \left(\frac{\mu^2}{\Lambda^2}\right)^{1-w}\right]^{-1} \xrightarrow{\Lambda\to\infty}\frac{16\pi^2}{2a}\frac{\alpha}{\alpha_c} \equiv y_*^2,
\label{running:y:b0}\\
        &\frac{\lambda_\Lambda(t)}{y_\Lambda^4(t)} = \frac{2a}{16\pi^2}\frac{\alpha_c}{\alpha}\left[1 - \left(\frac{\mu^2}{\Lambda^2}\right)^{2-2w}\right] \xrightarrow{\Lambda\to\infty} \frac{2a}{16\pi^2}\frac{\alpha_c}{\alpha} \equiv \frac{\lambda_*}{y_*^4},
\label{running:lambda:b0}
\end{align}
where $w$ is defined by $w\equiv1-\alpha/(2\alpha_c)$ with $\alpha_c\equiv 2\pi/c$. We also define $y_*$ and $\lambda_*$ at the large $\Lambda$ limit.
At this limit we can explicitly solve the RG equation (\ref{eq:RG}) for $m^2(t)$ and find
 \begin{align}
        m^2(t) = \frac{2a}{16\pi^2}\left(\frac{\Lambda^2}{\mu^2}\right)^w y_\Lambda^2(t)\mu^2\left(\frac{1}{g_4(\Lambda)} - \frac{1}{w}\right)
         \xrightarrow{\Lambda\to\infty} \frac{2a}{16\pi^2}y_*^2\mu^2 \left(\frac{1}{g_{4R}(\mu)} - \frac{1}{g_{4R}^*}\right) ,
\label{running:mass}
\end{align}
where $g_{4R}(\mu)$ denotes the renormalized four-fermion coupling and $g_{4R}^*$ is a constant parameter, see Ref.~\cite{Harada:1994wy}.
Below we consider only the fixed gauge coupling limit, $b=0$, for simplicity.

let us consider CMB fluctuations in the gauged NJL model. For this purpose,
we evaluate the RG improved effective potential for the composite field $\sigma$ in curved spacetime. We follow the procedure developed in Ref.~\cite{Geyer:1996wg}. The one-loop effective potential for the theory (\ref{L:gHY}) up to terms linear in $R$ is given by
\begin{align}
        V^{1 loop}(\sigma) = \frac{1}{2}m^2\sigma^2 + \frac{1}{4}\lambda\sigma^4 + \frac{1}{2}\xi R\sigma^2 - \frac{ay^4\sigma^4}{2\cdot 16\pi^2}\left(\ln\frac{y^2\sigma^2}{\mu^2} - \frac{3}{2}\right) - \frac{aRy^2\sigma^2}{12\cdot 16\pi^2}\left(\ln\frac{y^2\sigma^2}{\mu^2} - 1\right).
\label{effpot:oneloop}
\end{align}
To obtain the RG invariant effective potential of the gauged NJL model one introduces the compositeness condition into the effective potential (\ref{effpot:oneloop}) through the RG equations. The solution of the RG equation for the effective potential satisfies
\begin{align}
        V(g,y,\lambda,m^2,\xi,\sigma,\mu) = V(\bar{g}(t),\bar{y}(t),\bar{\lambda}(t),\bar{m}^2(t),\bar{\xi}(t),\bar{\sigma}(t),\mu e^t),
\label{RG:effpot}
\end{align}
where the barred quantities are the renormalized ones at the scale $\mu e^t$. The scale is fixed to drop the logarithmic terms in the RG invariant effective potential \cite{Harada:1994wy},
\begin{align}
        e^t=\left(\frac{y\sigma}{\mu}\right)^{1/(2-w)}.
        \label{scale:mu}
\end{align}
It should be noted that the gravitational couplings are not represented in Eq.~(\ref{RG:effpot}). These terms do not contribute at the linear curvature approximation.

Using the one-loop effective potential (\ref{effpot:oneloop}) as boundary condition, we calculate the RG invariant effective potential (\ref{RG:effpot}) from Eqs.~(\ref{running:y:b0}), (\ref{running:lambda:b0}), (\ref{running:mass}) and (\ref{scale:mu}).
At the limit $\Lambda\to\infty$ the RG invariant effective potential is given by
\begin{align}
        V(\sigma,\mu) = \frac{B(\mu)}{2}\sigma^2(\mu)+ \frac{C(\mu)}{4}[\sigma^2(\mu)]^{2/(2-w)}
        + \frac{D(\mu) R}{12}[\sigma^2(\mu)]^{1/(2-w)},
        \label{eq:effpot}
\end{align}
with
\begin{align}
  &B(\mu)\equiv
  \frac{\alpha}{\alpha_c}\left(\frac{1}{g_{4R}(\mu)} - \frac{1}{g_{4R}^*}\right)\mu^2,
\label{def:B}\\
  &C(\mu)\equiv
  \frac{2a}{16\pi^2}\left(\frac{16\pi^2}{2a\mu^2}\frac{\alpha}{\alpha_c}\right)^{2/(2-\omega)}
  \left(\frac{3}{2}+\frac{\alpha_c}{\alpha}\right)\mu^4,
\label{def:C}\\
  &D(\mu)\equiv
  \frac{2a}{16\pi^2}\left(\frac{16\pi^2}{2a\mu^2}\frac{\alpha}{\alpha_c}\right)^{1/(2-\omega)}
  \left(\frac{1}{2}+\frac{\alpha_c}{\alpha}\right)\mu^2.
\label{def:D}
\end{align}
Thus,  the gravitational effective action for the composite scalar field with weak curvature $R\ll\sigma^2$ in the gauged NJL model is obtained,
\begin{align}
        S = \int d^4x \sqrt{-g} \left[ - \frac{1}{2}R + \frac{1}{2}g^{\mu\nu}\pd_\mu\sigma\pd_\nu\sigma - V(\sigma,\mu)\right] ,
\label{eq:actionJ}
\end{align}
where the reduced Planck mass is set as $M_p=(8\pi G)^{-1/2}=1$.

\section{Inflationary parameters in gauged NJL model\label{InflationaryParameters}}
The gauged NJL model may give us an alternative scenario to induce the exponential expansion of our universe. Here we assume that the typical scale of the inflation is much smaller than the compositeness scale, $\Lambda$, and consider the effective potential (\ref{eq:effpot}) for the composite scalar field, $\sigma$, as an inflaton potential. Then we calculate the inflationary parameters generated by the composite scalar field.

For practical calculations it is more convenient to change the frame into the Einstein frame where the interaction term between the composite scalar and the curvature disappears. This change of the frame is achieved by the Weyl transformation,
\begin{equation}
\tilde{g}^{\mu\nu} = \Omega (x)^{-2} g^{\mu\nu},
\label{Weyl}
\end{equation}
where $\tilde{g}^{\mu\nu}$ is the metric tensor in the transformed frame and the Weyl factor, $\Omega(x)$, is an analytic function with respect to the space-time coordinates.
Applying the Weyl transformation (\ref{Weyl}) and setting the conformal factor to be
\begin{align}
\Omega^2 = 1+ \frac{D(\mu)}{6}(\sigma^2)^{1/(2-w)},
\label{omega}
\end{align}
one gets the effective action (\ref{eq:actionJ}) in the Einstein frame
\begin{align}
        S \rightarrow \int d^4x\sqrt{-\tilde{g}}\left[ - \frac{1}{2}\tilde{R} + \frac{3}{4}\Omega^{-4}\tilde{g}^{\mu\nu}\pd_\mu\Omega^2\pd_\nu\Omega^2 + \Omega^{-4}\left(\frac{1}{2}g^{\mu\nu}\pd_\mu\sigma\pd_\nu\sigma - \frac{B(\mu)}{2}\sigma^2 - \frac{C(\mu)}{4}(\sigma^2)^{2/(2-w)}\right)\right],
\label{eq:actionE}
\end{align}
where $\tilde{g}$ and $\tilde{R}$ are the determinant of the metric tensor and the Ricci scalar in the transformed frame, respectively.
In order to obtain the canonical kinetic term we redefine the auxiliary field $\sigma$ to satisfy the relation,
\begin{align}
        d\varphi \equiv \left[\frac{3}{2}\left(\frac{1}{\Omega^2}\frac{2}{2-w}\frac{D(\mu)}{6}(\sigma^2)^{w/(4-2w)}\right)^2 + \frac{1}{\Omega^2}\right]^{1/2} d\sigma
\end{align}
Then the effective action (\ref{eq:actionE}) reduces to
\begin{align}
S_E = \int d^4x \sqrt{-\tilde{g}}\left( -  \frac{1}{2}\tilde{R} + \frac{1}{2}\tilde{g}^{\mu\nu}\pd_\mu\varphi\pd_\nu\varphi - V_E(\sigma,\mu)\right),
\label{act:se}
\end{align}
where the subscript $E$ denotes the Einstein frame and the effective potential $V_E(\sigma,\mu)$ is defined by
\begin{align}
V_E(\sigma,\mu) \equiv \Omega^{-4}\left[\frac{B(\mu)}{2}\sigma^2 + \frac{C(\mu)}{4}(\sigma^2)^{2/(2 - w)}\right].
\label{pot:ve}
\end{align}
The effective potential $V_E(\sigma,\mu)$ in the Einstein frame depends on the parameters in the original gauged NJL model through Eqs.(\ref{def:B}) - (\ref{def:D}). The quantum effects are encoded via the $\mu$-dependence of these parameters with the compositeness condition. It should be noted that the obtained inflationary parameters are just the same in both the original and Einstein frames \cite{Kaiser:1994}.

We will not be interesting in the exact solutions for the action (\ref{act:se}) as it has been discussed in detail for various choices of the potential. We will concentrate here on quasi-de Sitter phase which eventually describes the inflationary dynamics. In the study of quasi-de Sitter phase we apply the standard slow-roll approximation which is well-known (see Refs.~\cite{Linde:2005ht}, for instance) to be consistent approximation to describe the quasi-de Sitter phase.

We consider a locally flat Friedman-Robertson-Walker (FRW) universe defined by the metric,
\begin{align}
  ds^2=dt^2-a^2(t)(dx^2+dy^2+dz^2),
\end{align}
where the evolution of the spacetime is described by the temporal development of the scale factor, $a(t)$.
In the FRW universe, the equation of motion for $\varphi$ and the Freedman equation are given by
\begin{align}
  \ddot{\varphi}+3H \dot{\varphi} = - \frac{\pd V_E}{\pd \varphi} ,
\label{eom}
\end{align}
\begin{align}
  H^2 = \frac{1}{3} \left(\frac{1}{2}\dot{\varphi}^2+V_E\right) ,
\label{freedman}
\end{align}
with the Hubble parameter, $H\equiv \dot{a}/a$, respectively. In the slow roll scenario we assume that the spacetime is slowly varying and impose the conditions
\begin{align}
  \dot{\varphi}^2 \ll V,\,\, \ddot{\varphi} \ll \left|\frac{\pd V_E}{\pd \varphi}\right| .
\label{slowroll}
\end{align}
Then we obtain the quasi-de Sitter phase,
\begin{align}
  a(t+\Delta t)\sim x a(t)e^{\sqrt{\frac{V}{3}}\Delta t}+ (1-x)a(t)e^{-\sqrt{\frac{V}{3}}\Delta t} ,
\label{deSitter}
\end{align}
where $x$ is a constant parameter.

In a standard scenario of the the inflation the CMB fluctuation arises from the primordial quantum fluctuation for the inflaton field. Here we employ the slow-roll approximation for the inflationary expansion of the universe. Under the approximation the inflationary parameters can be fully represented by means of the inflaton potential,
\begin{align}
        &\epsilon = \frac{1}{2} \left(\frac{1}{V_E}\frac{\pd V_E}{\pd \varphi}\right)^2
        =  \frac{1}{2} \left(\frac{1}{V_E}\frac{\pd V_E}{\pd \sigma} \frac{\pd \sigma}{\pd \varphi}\right)^2, \\
        &\eta = \frac{1}{V_E}\frac{\pd^2 V_E}{\pd \varphi^2}
        = \frac{1}{V_E}\left[\frac{\pd }{\pd \sigma}\left(\frac{\pd V_E}{\pd \sigma} \frac{\pd \sigma}{\pd \varphi}\right)\right]
        \frac{\pd \sigma}{\pd \varphi}, \\
        &\xi =  \frac{1}{V_E^2}\frac{\pd V_E}{\pd \varphi}\frac{\pd^3 V_E}{\pd \varphi^3}
        =\frac{1}{V_E^2}\frac{\pd V_E}{\pd \sigma}\frac{\pd \sigma}{\pd \varphi}
        \left\{\frac{\pd }{\pd \sigma}\left[\frac{\pd }{\pd \sigma} \left(\frac{\pd V_E}{\pd \sigma} \frac{\pd \sigma}{\pd \varphi}\right)
        \frac{\pd \sigma}{\pd \varphi}\right]\right\}\frac{\pd \sigma}{\pd \varphi} .
\end{align}
The e-folding number $N$ is given by
\begin{align}
        N = \int\frac{V_E}{\pd V_E/\pd\varphi}d\varphi
        =\int^{\sigma_N}_{\sigma_{end}}\left(\frac{\pd \sigma}{\pd \varphi}\right)^2\frac{V_E}{\pd V_E/\pd\sigma}d\sigma .
\label{efold}
\end{align}
where the integration is performed on the interval from the horizon crossing $\sigma=\sigma_N$ to the end of the inflation $\sigma=\sigma_{end}$. Since the inflationary parameters have to be small, the field configuration, $\sigma_{end}$, can be estimated where one of the inflationary parameters exceeds order one. We noted that the change of the value for $\sigma_{end}$ gives only a small correction to the e-folding number $N$. The dominant contribution comes from the vale at  the horizon crossing, $\sigma_{N}$, where the derivative of the effective potential is small enough. In the other words the field configuration, $\sigma_{N}$ is obtained as a function of the e-folding number $N$ from Eq.(\ref{efold}). Then one can evaluate the density fluctuation, $\delta$, by \cite{Linde:2005ht}
\begin{align}
        \delta \sim \left.\frac{V_E^{3/2}}{\sqrt{12\pi^2}}\left(\frac{\pd V_E}{\pd \sigma}\frac{\pd \sigma}{\pd \varphi}\right)^{-1}\right|_{\sigma=\sigma_N} .
\end{align}
The spectral index, $n_s$, the tensor-to-scalar ratio, $r$, and the running of the spectral index, $\alpha_s$, are represented by using the slow-roll parameters \cite{Kohri:2013mxa},
\begin{align}
        &n_s -1 = 2\left.\eta\right|_{\sigma=\sigma_N} - 6\left. \epsilon\right|_{\sigma=\sigma_N}  ,
\label{ns}\\
        &r = 16\left. \epsilon\right|_{\sigma=\sigma_N}  ,
\label{r}\\
        &\alpha_s = \frac{dn_s}{d \ln k} = -24 \left. \epsilon^2\right|_{\sigma=\sigma_N} + 16 \left. \epsilon\eta\right|_{\sigma=\sigma_N}
        - 2 \left. \xi \right|_{\sigma=\sigma_N} ,
\label{running:ns}
\end{align}
where we omit higher-order terms. Thus the observed CMB constraints for $\delta$, $n_s$, $r$ and $\alpha_s$ restrict the possible range of the parameters in the gauged NJL model for quasi-de Sitter universe under discussion..

\section{Numerical results\label{NumericalResults}}
The effective action $(\ref{eq:actionJ})$ depends on the fine-structure constant $\alpha$, the renormalized four-fermion coupling $g_{4R}$, the number of fermion species, $N_f$, $N_c$ and the renormalization scale $\mu$.
In this section we numerically calculate the inflationary parameters $\delta$, $n_s$, $r$ and $\alpha_s$ with varying the model parameters, $N_f$, $N_c$, $g_{4R}$, $\alpha$, $\mu$ and the e-folding number, $N$.

\subsection{Small number of fermion species ($N_f=1, \, N_c = 10$)}
\begin{figure}[htbp]
\begin{minipage}{0.48\hsize}
        \centering
        \includegraphics[width=0.84\linewidth]{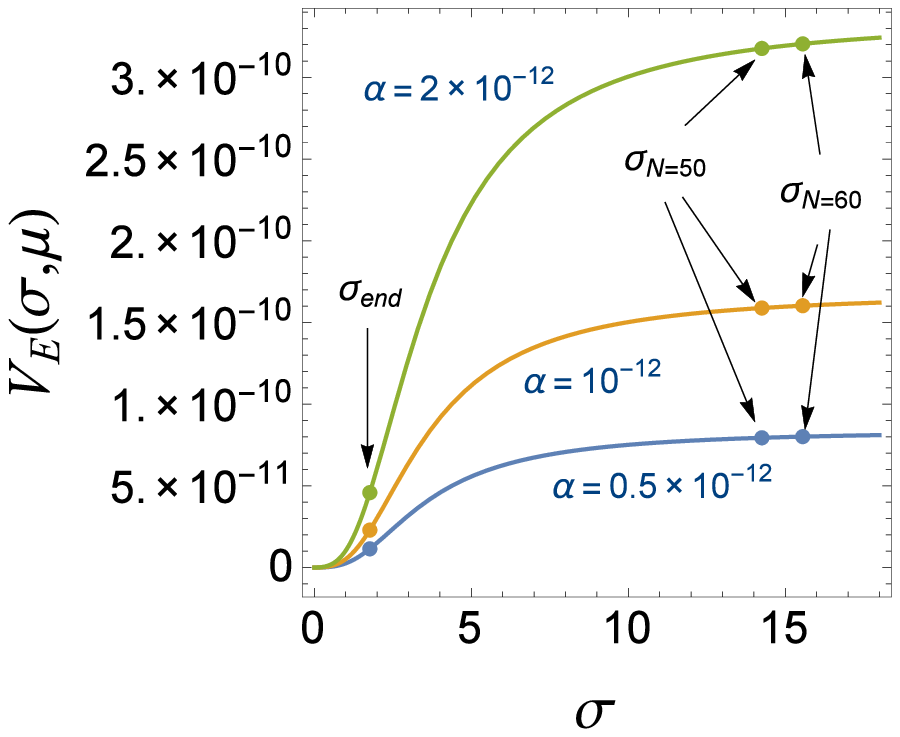}
        \vglue -4mm
        \caption{Behavior of the potential $V_E(\sigma, \mu)$ as a function of $\sigma$ for $N_f = 1$, $N_c = 10$, $1/g_{4R}-1/g_{4R}^* = 1$, $\mu=1$ and $\alpha=0.5\times 10^{-12}, 1\times 10^{-12}, 2\times 10^{-12}$.}
        \label{fig:potential}
\end{minipage}
\hspace{0.04\hsize}
\begin{minipage}{0.48\hsize}
        \centering
        \includegraphics[width=0.84\linewidth]{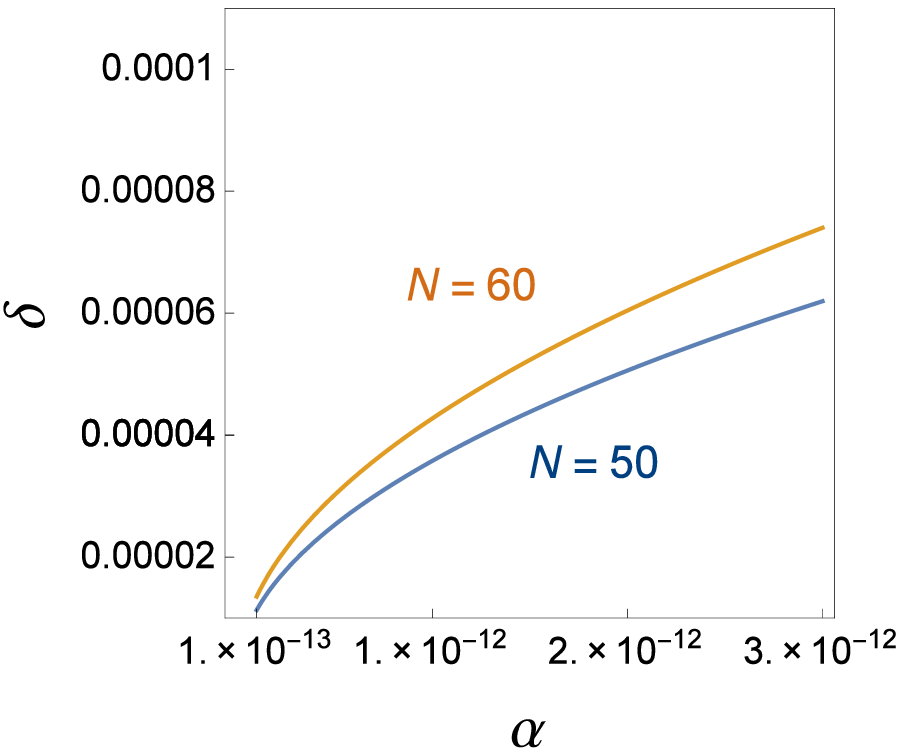}
        \vglue -4mm
        \caption{Behavior of the density fluctuation $\delta$ as a function of $\alpha$ for $N_f = 1$, $N_c = 10$, $1/g_{4R}-1/g_{4R}^* = 1$ and $\mu=1$.}
        \label{fig:delta}
\end{minipage}
\end{figure}

In this subsection we consider a model with a small number of fermion species, $N_f=1, \, N_c = 10$.
First we set model parameters as
$1/g_{4R}-1/g_{4R}^* = 1,\, \mu =1,$
and study the gauge coupling dependence. The behavior of the effective potential is drawn in Fig.~\ref{fig:potential} for $\alpha=0.5\times 10^{-12}, 1\times 10^{-12}$ and $2\times 10^{-12}$. We plot the points, $\sigma=\sigma_{end}$, $\sigma=\sigma_{N=50}$ and $\sigma=\sigma_{N=60}$ on each lines. A gentler slope of the potential is observed for a weaker gauge coupling. It should be noticed that we have to fine-tune the field configuration at the horizon crossing to obtain the enough e-foldings in the case of the NJL model (See Appendix A).
The density fluctuation decreases as the slope becomes gentle. As is shown in Fig.~\ref{fig:delta}, the density fluctuation is monotonically increasing as a function of $\alpha$. The observed density fluctuation $\delta \sim 4.93\times10^{-5}$ is obtained for $\alpha \sim O(10^{-12})$.

In Table~\ref{tab:alpha}  the values for the inflationary parameters $n_s$, $r$ and $\alpha_s$ for $\alpha = 10^{-12}, 10^{-11}$ and $10^{-10}$ are shown. The spectrum index $n_s$, tensor-scalar ratio $r$, and the running of the spectral index, $\alpha_s$, are consistent with the observed Planck constraints, $n_s = 0.9655\pm0.0062$, $\alpha_s =-0.0084\pm0.0082$ (68\% CL, PlanckTT+lowP) and $r_{0.002} < 0.10$ (95\% CL, PlanckTT+lowP)\cite{Ade:2015lrj}. It is also found that the $\alpha$-dependence for the inflationary parameters $n_s$, $r$ and $\alpha_s$ is negligible.

\begin{table}[htb]
\begin{center}
\begin{tabular}{cccccc}
\hline
$\alpha$ & $\delta$ & $n_s$ & $r$ & $\alpha_s$ \\
        $10^{-12}$ & $3.58\times 10^{-5}$ & 0.961 & 0.0084 & -0.00077 \\
        $10^{-11}$ & $1.13\times 10^{-4}$ & 0.961 & 0.0084 & -0.00077 \\
        $10^{-10}$ & $3.58\times 10^{-4}$ & 0.961 & 0.0084 & -0.00077 \\
\hline
\end{tabular}
\caption{Inflationary parameters for $N_f = 1$, $N_c = 10$, $1/g_{4R}-1/g_{4R}^* = 1$, $\mu=1$ and $N=50$.
\label{tab:alpha}}
\end{center}
\end{table}

Next we discuss the dependence on the renormalized four-fermion coupling, $g_{4R}$ and the renormalization scale, $\mu$.  Since a realistic value for the density fluctuation is obtained for $\alpha \sim O(10^{-12})$, we set $\alpha = 10^{-12}$ and numerically calculate the inflationary parameters as $\mu$ and $g_{4R}$ vary. We show some typical values in Table~\ref{tab:mug4}. The $\mu$ and $g_{4R}$ dependences are also negligible for the density fluctuation, $\delta$, the spectral index, $n_s$, the tensor-to-scalar ratio, $r$, and the running of the spectral index, $\alpha_s$.

\begin{table}[htb]
\begin{center}
\begin{tabular}{cccccc}
\hline
$\mu$ & $1/g_{4R}-1/g_{4R}^*$ & $\delta$ & $n_s$ & $r$ & $\alpha_s$ \\
        $10^{-8}$ & 1 &  $3.59\times 10^{-5}$ & 0.961 & 0.0083 & -0.00076 \\
        $10^{-4}$ & 1 & $3.59\times 10^{-5}$ & 0.961 & 0.0083 & -0.00076 \\
        1 & 1 & $3.58\times 10^{-5}$ & 0.961 & 0.0084 & -0.00077 \\
        1 &  $10^{-4}$ & $3.59\times 10^{-5}$ & 0.961 & 0.0083 & -0.00076 \\
        1 & $10^{-8}$ & $3.59\times 10^{-5}$ & 0.961 & 0.0083 & -0.00076 \\
\hline
\end{tabular}
\caption{Inflationary parameters for $N_f = 1$, $N_c = 10$, $\alpha=10^{-12}$ and $N=50$.
\label{tab:mug4}}
\end{center}
\end{table}

It is found that the spectral index, the tensor-to-scalar ratio and the running of the spectral index are almost fixed at $n_s=0.961$, $r=0.0083$, $\alpha_s=-0.00076$ for $N=50$ in $N_f=1$, $N_c = 10$ case. These values are consistent with the Planck 2015 data.

\subsection{Large number of fermion species}

Let us investigate the inflationary parameters at  fixed gauge coupling limit, $b\rightarrow +0$ for simplicity. The assumption is valid for $11N_c\sim 2N_f$. However, we vary the number of the fermion species, $N_f$ and $N_c$ , with a fixed gauge coupling, $g$. Since  the extremely small gauge coupling,  $\alpha \sim O(10^{-12})$, is considered the running of the gauge coupling in Eq.~(\ref{eq:RG}) can be neglected. The effective potential develops a non-trivial extremum for $N_f N_c > 24 \pi^2$ at the small gauge coupling limit. It induces the fine-tuning problem discussed in App. \ref{NJL}. In order to avoid the problem we consider the limited case $N_f N_c < 24 \pi^2$.

\begin{figure}[htbp]
\begin{minipage}{0.48\hsize}
        \centering
        \includegraphics[width=0.9\linewidth]{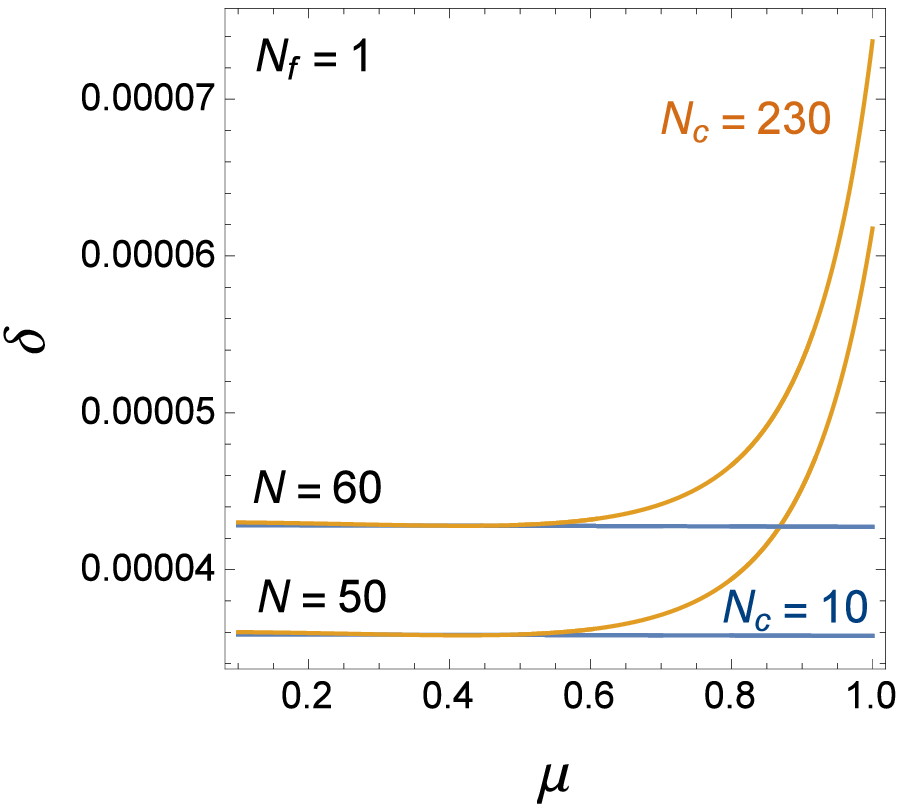}
        \vglue -4mm
        \caption{Behavior of  the density fluctuation $\delta$ as a function of $\mu$ for $N_f=1$, $\alpha=10^{-12}$, $1/g_{4R}-1/g_{4R}^* = 1$.}
\label{fig:deltamu:Nc}
\end{minipage}
\hspace{0.04\hsize}
\begin{minipage}{0.48\hsize}
        \centering
        \includegraphics[width=0.92\linewidth]{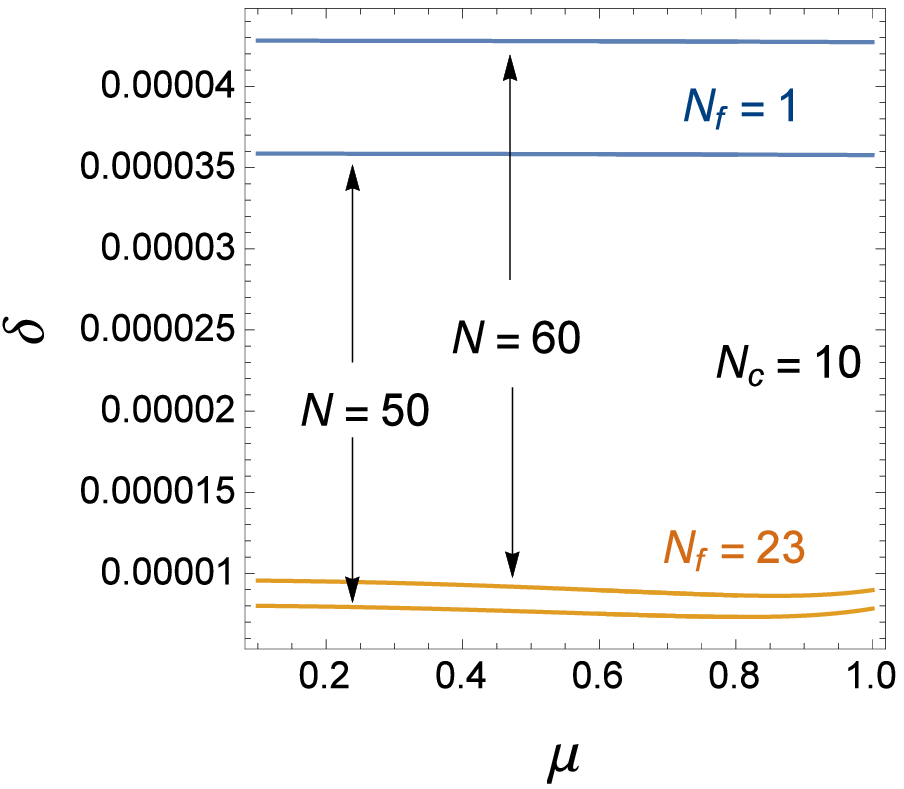}
        \vglue -4mm
        \caption{Behavior of  the density fluctuation $\delta$ as a function of $\mu$ for $N_c=10$, $\alpha=10^{-12}$, $1/g_{4R}-1/g_{4R}^* = 1$.}
\label{fig:deltamu:Nf}
\end{minipage}
\end{figure}

In Figs.~\ref{fig:deltamu:Nc} and \ref{fig:deltamu:Nf} we compare the behavior of the density fluctuation between a small and a large number of fermion species. The density fluctuation grows up near $\mu=1$ for $\{N_f, N_c\}=\{1, 230\}$ and $\{10, 23\}$ cases. Non-negligible $\mu$ dependence appears for a larger $N_f$ or $N_c$ cases. As is shown in Fig.~\ref{fig:deltamu:Nc} the obtained density fluctuation for $N_c=230$ approaches to the one for $N_c=10$ as the renormalization scale $\mu$ decreases. We note that lines for $N_c=10$ and $N_c=230$ cross twice in $\mu\in[0,1]$ and a small but a finite difference remains at $\mu\rightarrow 0$ limit. In Fig.~\ref{fig:deltamu:Nf} lines for $N_f=1$ and $N_f=23$ do not cross in $\mu\in[0,1]$ and a larger difference observed at the limit, $\mu\rightarrow 0$.

\begin{figure}[htbp]
\begin{minipage}{0.48\hsize}
        \centering
        \includegraphics[width=0.85\linewidth]{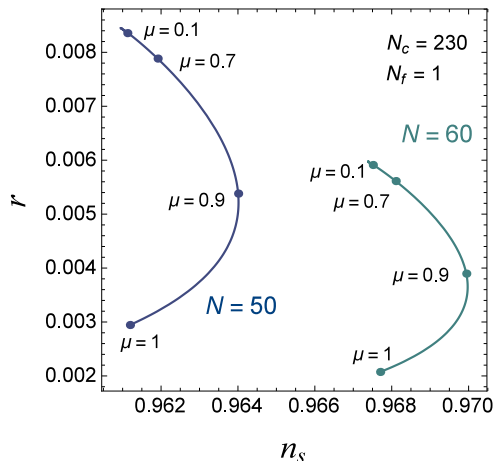}
\end{minipage}
\hspace{0.04\hsize}
\begin{minipage}{0.48\hsize}
        \centering
        \includegraphics[width=0.94\linewidth]{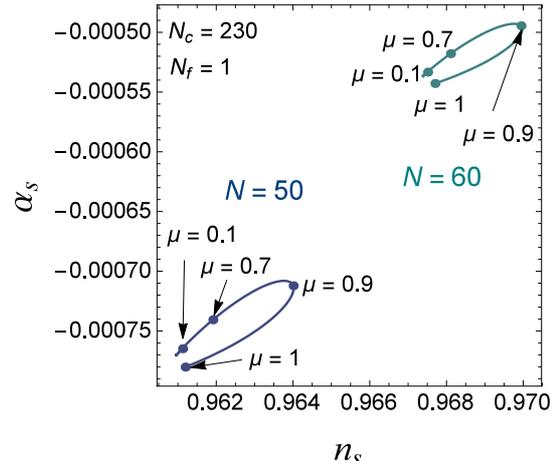}
\end{minipage}
\vglue -4mm
\caption{Behavior of  the tensor-to-scalar ratio $r$ and the running of the spectral index $\alpha_s$ as a function of  the spectral index $n_s$ for $\alpha=10^{-12}$, $N_f = 1$, $N_c=230$ and $1/g_{4R}-1/g_{4R}^* = 1$.
}
\label{fig:ralpsmu:Nc}
\end{figure}

\begin{figure}[htbp]
\begin{minipage}{0.48\hsize}
        \centering
        \includegraphics[width=0.85\linewidth]{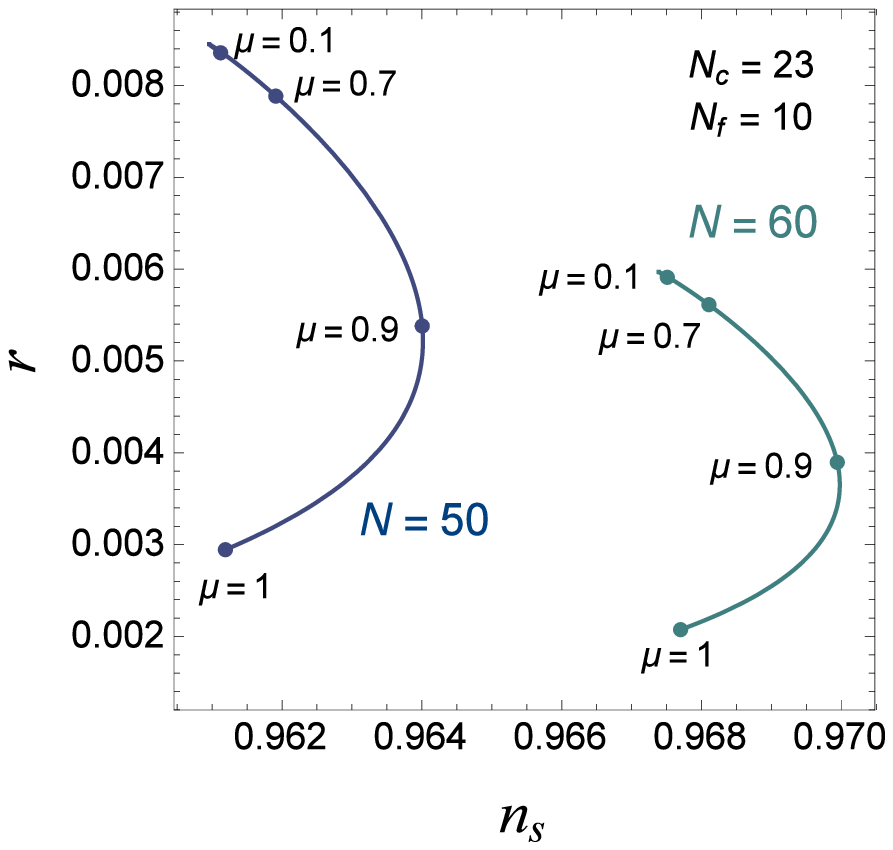}
\end{minipage}
\hspace{0.04\hsize}
\begin{minipage}{0.48\hsize}
        \centering
        \includegraphics[width=0.94\linewidth]{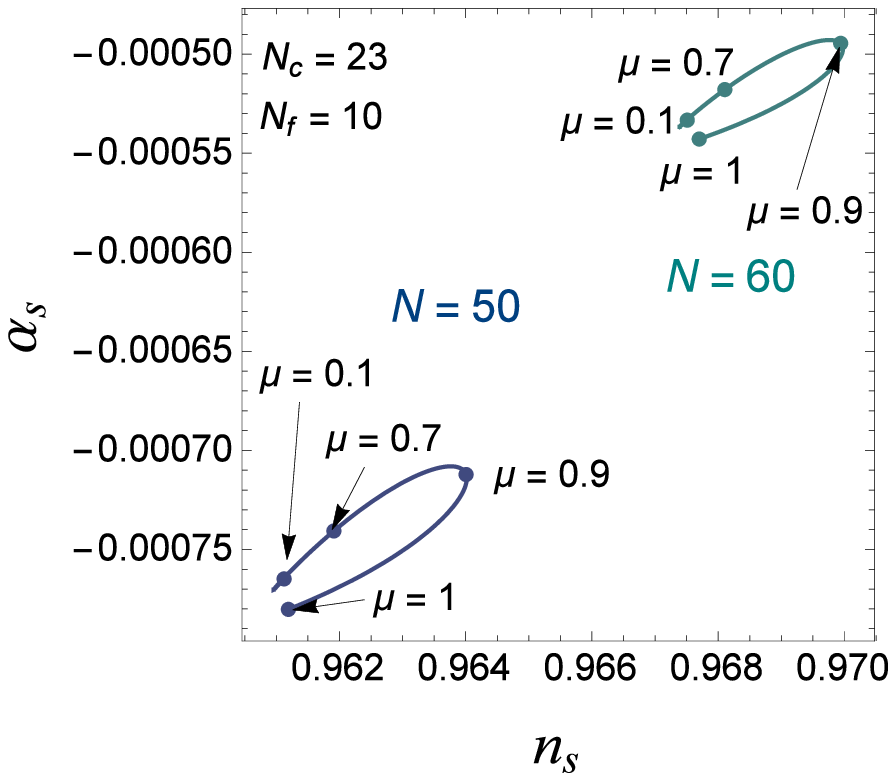}
\end{minipage}
\vglue -4mm
\caption{Behavior of  the tensor-to-scalar ratio $r$ and the running of the spectral index $\alpha_s$ as a function of  the spectral index $n_s$ for $\alpha=10^{-12}$, $N_f = 10$, $N_c=23$ and $1/g_{4R}-1/g_{4R}^* = 1$.
}
\label{fig:ralpsmu:Nf}
\end{figure}

For a larger number of fermion species the $\mu$ dependence appears in the spectral index, $n_s$, the tensor-to-scalar ratio, $r$, and the running of the spectral index, $\alpha_s$ in models. We draw the behavior of these inflationary parameters on the interval $\mu= [0.1,1.0]$ in Figs.~\ref{fig:ralpsmu:Nc} and \ref{fig:ralpsmu:Nf}. It should be noticed that we normalize  the mass scale by the Planck scale and consider the large $\Lambda$ limit. Such assumption is suitable for $\mu \ll 1.0\sim M_p$. A similar behavior is observed for $\{N_f, N_c\}=\{1, 230\}$ and $\{10, 23\}$ cases. Thus $n_s$, $r$ and $\alpha_s$ seem to  depend on the product $N_f N_c$. At the limit, $\mu\rightarrow 0$, the result approaches to the one for a small number of fermion species, $N_f=1$ and $N_c=23$. The obtained trajectories are shifted as the e-folding number varies. All the trajectories are consistent with the Planck 2015 observations.

\begin{figure}[htbp]
\begin{minipage}{0.48\hsize}
        \centering
        \includegraphics[width=0.9\linewidth]{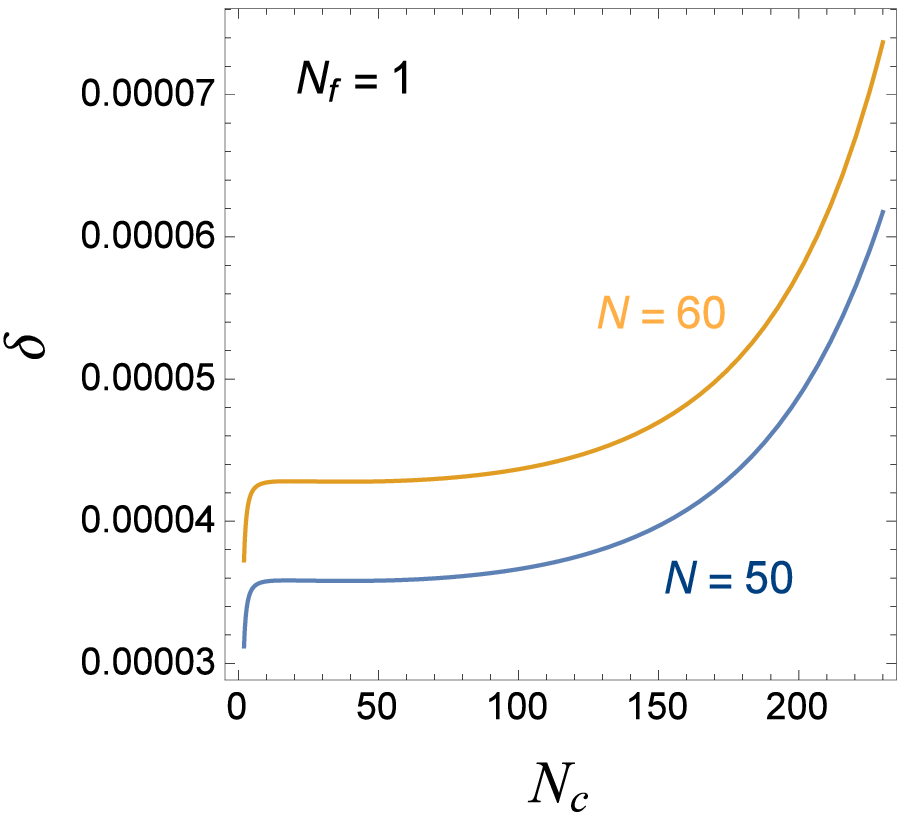}
        \vglue -4mm
        \caption{Behavior of  the density fluctuation $\delta$ as a function of $N_c$ for $N_f=1$, $\alpha=10^{-12}$, $1/g_{4R}-1/g_{4R}^* = 1$ and $\mu=1$.}
        \label{fig:delta:Nc}
\end{minipage}
\hspace{0.04\hsize}
\begin{minipage}{0.48\hsize}
        \centering
        \includegraphics[width=0.92\linewidth]{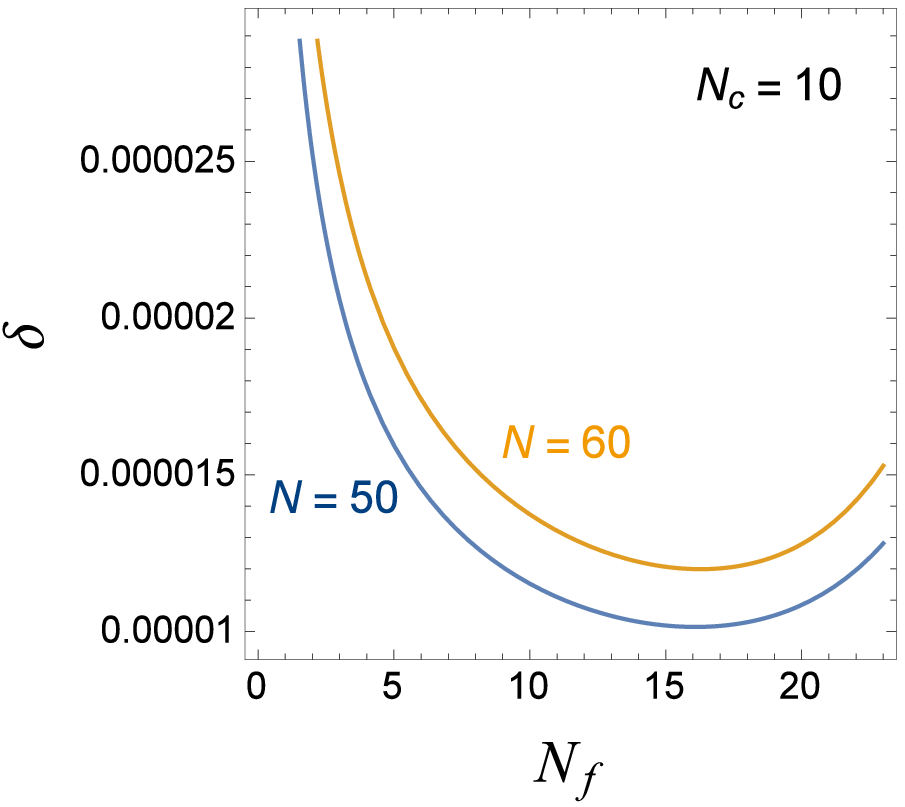}
        \vglue -4mm
        \caption{Behavior of  the density fluctuation $\delta$ as a function of $N_f$ for $N_c=10$, $\alpha=10^{-12}$, $1/g_{4R}-1/g_{4R}^* = 1$ and $\mu=1$.}
\label{fig:delta:Nf}
\end{minipage}
\end{figure}

 The inflationary parameters depend on the renormalization scale $\mu$ for a larger number of the fermion species. It is observed that the parameters, $n_s$, $r$ and $\alpha_s$ simply approach to the values for a small number of the fermion species at the small $\mu$ limit.
Next we fix the model parameters at $\mu=1$, $\alpha=10^{-12}$, $1/g_{4R}-1/g_{4R}^* = 1$ and evaluate the $N_f$ and $N_c$ dependences. The behavior of the density fluctuation, $\delta$, is plotted as $N_c$ and $N_f$ vary in Figs.~\ref{fig:delta:Nc} and \ref{fig:delta:Nf}, respectively. The range of the density fluctuation is smaller than the one in Fig.~\ref{fig:delta}. The number of the fermion species induces a sub-dominant contribution to the density fluctuation compared with the gauge coupling. Thus, we can obtain a suitable density fluctuation by tuning the strength of the gauge coupling.

\begin{figure}[ht]
\begin{minipage}{0.48\hsize}
        \centering
        \includegraphics[width=0.85\linewidth]{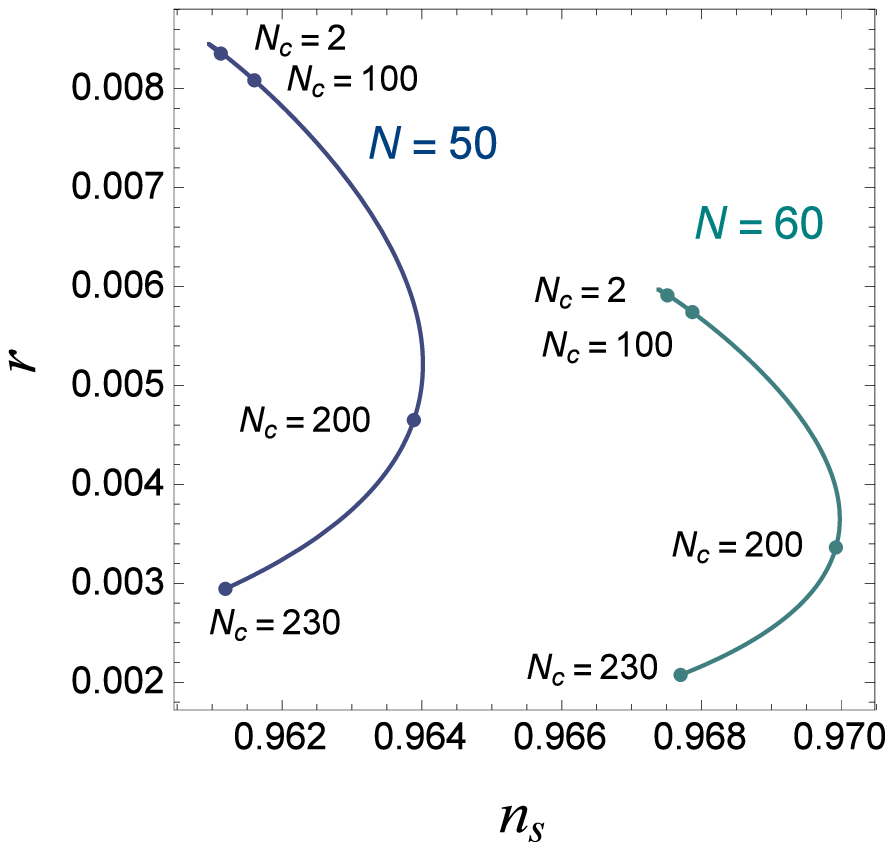}
\end{minipage}
\hspace{0.04\hsize}
\begin{minipage}{0.48\hsize}
        \centering
        \includegraphics[width=0.94\linewidth]{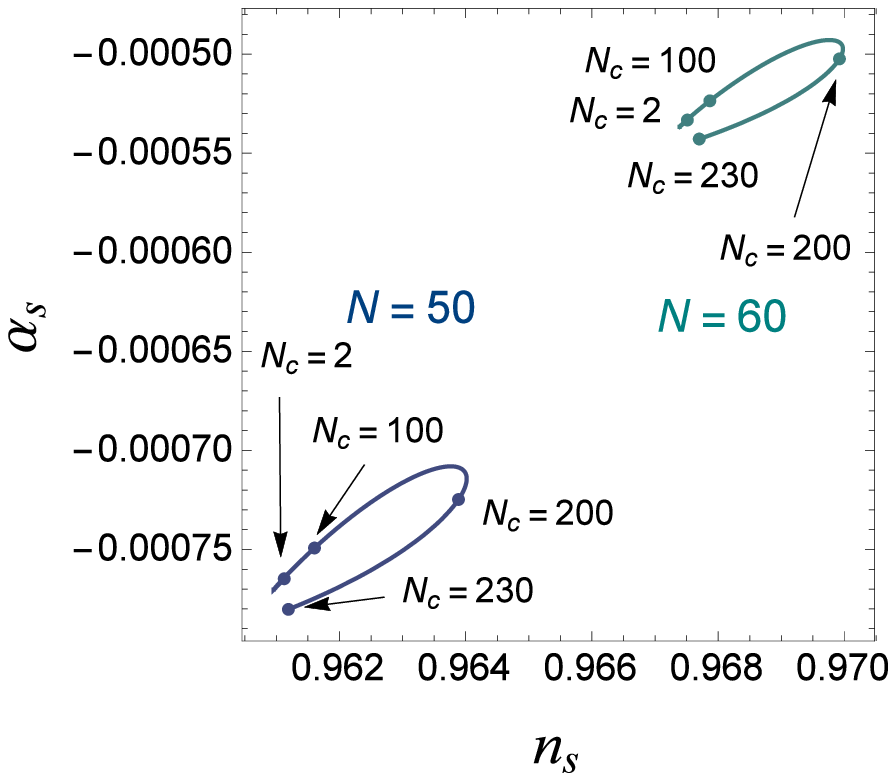}
\end{minipage}
\vglue -4mm
\caption{Behavior of  the tensor-to-scalar ratio $r$ and the running of the spectral index $\alpha_s$ as a function of  the spectral index $n_s$ for $\alpha=10^{-12}$, $N_f = 1$, $1/g_{4R}-1/g_{4R}^* = 1$ and $\mu=1$.}
\label{fig:ralps:Nc}
\end{figure}

\begin{figure}[ht]
\begin{minipage}{0.48\hsize}
        \centering
        \includegraphics[width=0.85\linewidth]{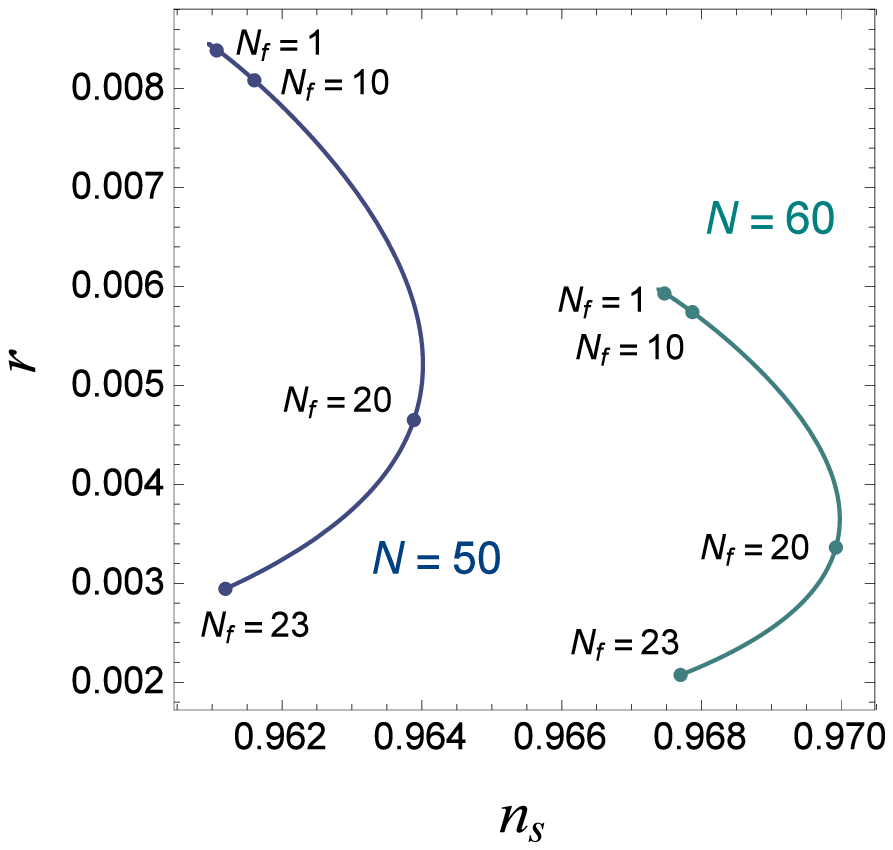}
\end{minipage}
\hspace{0.04\hsize}
\begin{minipage}{0.48\hsize}
        \centering
        \includegraphics[width=0.94\linewidth]{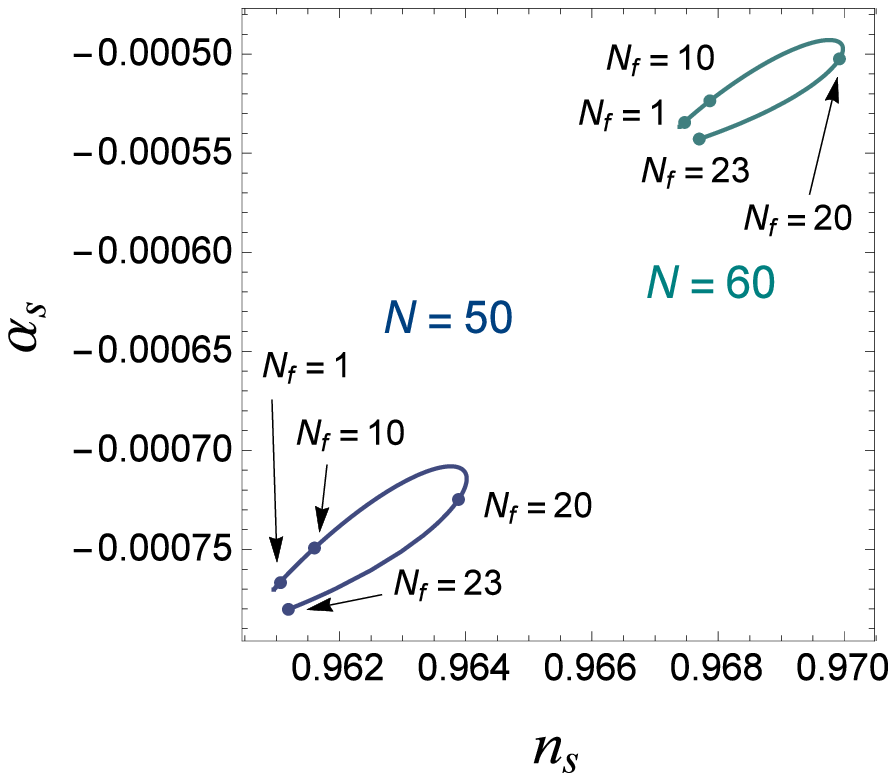}
\end{minipage}
\vglue -4mm
\caption{Behavior of  the tensor-to-scalar ratio $r$ and the running of the spectral index $\alpha_s$ as a function of  the spectral index $n_s$ for $\alpha=10^{-12}$, $N_c = 10$, $1/g_{4R}-1/g_{4R}^* = 1$ and $\mu=1$.}
\label{fig:ralps:Nf}
\end{figure}

In Figs.~\ref{fig:ralps:Nc} and \ref{fig:ralps:Nf} the tensor-to-scalar ratio, $r$, and the running of the spectral index, $\alpha_s$, are plotted as functions of  the spectral index, $n_s$, for the e-folding number, $N=50$ and $60$. Both the figures show a similar behavior to Figs.~\ref{fig:ralpsmu:Nc} and \ref{fig:ralpsmu:Nf}. A model with a large number of fermion species at $\mu\sim 1$ and a small number of fermion species at a small $\mu$ generate nearly equal values of $n_s$, $r$ and $\alpha_s$. The obtained trajectories are consistent
 with the Planck 2015 results. The tensor-to-scalar ratio, $r$, monotonically decreases as $N_c$ and $N_f$ increases. The spectral index, $n_s$ and the running of the spectral index, $\alpha_s$ have maximal values at $a= 2N_f N_c \sim 380$.

Thus, we demonstrated that realistic gauged NJL-model under discussion for specific values of parameters is consistent with Planck cosmological data. From another side, the model under discussion is realistic high energy physics model of elementary particles. For numerical calculations, we considered several choices of fermion species so that such choice is quite standard in particle physics. Thus, we arrive to realistic (composite) particle physics model which predicts viable inflation.

\section{Exit from inflation\label{EndInflation}}
The end of the slow roll era is found by observing the parameters, $\epsilon$ and $\eta$. In the present model the parameter $\epsilon$ exceeds unity at $\sigma=\sigma_{end}$ in Fig.~\ref{fig:potential}. Under the slow roll approximation, the deceleration parameter, $q$, is written as
\begin{align}
  q\equiv -\frac{\ddot{a}}{H^2 a} \sim \epsilon -1 .
\end{align}
The deceleration parameter changes the sign at $\sigma=\sigma_{end}$ and develops a positive value. Thus the universe turns to decelerated expansion era.

In this paper we set the initial condition $\sigma_N$ to generate the e-folding number between 50 and 60.
Thus the duration of the inflation can be estimated by
\begin{align}
t_{duration} = \int_{\sigma_{end}}^{\sigma_N} \left(\frac{\pd \sigma}{\pd \varphi}\right)^2\frac{\sqrt{3V_E}}{\pd V_E/\pd\sigma}d\sigma .
\end{align}
We numerically find
\begin{align}
t_{duration} = (7\sim 8)\times 10^6 M_p^{-1} \sim 10^{-36} \, \mbox{s}.
\end{align}

From Eqs.~(\ref{eom}) and (\ref{freedman}) we obtain
\begin{align}
H^2 =\frac{1}{3}(-\dot{H}+V_E) .
\label{dynamics}
\end{align}
We define the perturbation $\Delta H(t)$ by
\begin{align}
  H=H_{dS}+\Delta H(t) ,
\label{perturbation}
\end{align}
with
$H_{dS}\equiv V_E/3$. Inserting Eq.~(\ref{perturbation}) into Eq.~(\ref{dynamics}), we get
\begin{align}
  \frac{\Delta \dot{H}}{\Delta H} \sim -6 H_{dS} +\mbox{O}(\Delta H^2).
\label{dynamics:2}
\end{align}
The solution of this equation is given by
\begin{align}
  \Delta H(t) = A \sinh\left(6\sqrt{\frac{V}{3}}t \right),
\label{sol:dynamics}
\end{align}
where we set an initial condition $\Delta H(t=0)=0$ and $A$ is a constant parameter. Thus the de Sitter solution (\ref{deSitter}) is unstable.
As is shown in Fig.~\ref{fig:potential}, the potential energy extremely decreases at $\sigma=\sigma_{end}$. Then the dominant contribution to derive the expansion of the universe comes from the radiation. Therefore, the standard cosmology is recovered. It should be noted that $R^2$ terms induced by the trace anomaly also unstabilizes the de Sitter solution and then the radiation dominant era realizes \cite{Bamba:2014}.

\section{Conclusions\label{Conclusion}}
In summary, we proposed gauged NJL inflation as an alternative for Higgs inflation. Specifically, the gauged NJL model is studied as the realistic particle physics model for the composite scalar.
 Applying the auxiliary field method and performing the renormalization group improvement (for review of RG in curved spacetime, see \cite{buc}), the model is represented by the gauge-Higgs-Yukawa theory. We  assumed that the gauge-Higgs-Yukawa theory well describes the model at the scale of inflation at the $\Lambda\rightarrow\infty$ limit. Then the produced CMB fluctuations have been calculated in the gauge-Higgs-Yukawa theory with the compositeness conditions at the fixed gauge coupling.

The behavior for inflationary parameters $\delta$, $n_s$, $r$ and $\alpha_s$ has been numerically evaluated as a function of the couplings, the renormalization scale, the e-folding number and the number of the fermion species. It is found that the smallness of the density fluctuation can be realized by tuning the gauge coupling. We proved that the model with a small number of fermion species, $N_f=1, \, N_c = 10$ satisfies the Planck 2015 data. In this case the spectral index, the tensor-to-scalar ratio and the running of the spectral index develop almost fixed values, $n_s=0.961$, $r=0.0083$, $\alpha_s=-0.00076$ for $N=50$. For a large number of fermion species, these inflationary parameters depend on the renormalization scale $\mu$. However, the obtained trajectories of the inflationary parameters are consistent with the Planck 2015 data. Therefore,  the gauged NJL model naturally generates inflationary parameters consistent with the Planck 2015 observation and may serve as viable inflationary theory.

At the next step, one has to consider the end of inflation and graceful exit. It is clear that higher-derivative RG corrected terms (like $R^2$, etc) which are induced as vacuum polarization beyond the linear-curvature approximation may naturally provide the graceful exit.  It is also interesting to consider the reheating process in the gauged NJL model.
In fact, this goes beyond this work. However, the preliminary considerations of preheating process indicate reheating maybe successful as it is in usual scalar theories of inflation (in fact, our model is effectively scalar inflationary theory).
These questions will be considered elsewhere.
Furthermore,  NJL model may pretend to the role of dark matter (see, for instance, Ref.~\cite{Holthausen:2013ota}), hence presumably unifying inflation with dark matter.

\section*{Acknowledgements}
The authors are grateful to T.~Kugo for useful comments.
The work by TI is supported in part by JSPS KAKENHI Grant Number 26400250 and that by SDO is supported in part by MINECO (Spain), project FIS2013-44881 and by MES (Russia).

\appendix

\section{NJL limit\label{NJL}}
Let us discuss the slow-roll solution in the NJL model. It is obtained by omitting the gauge interaction. Here we consider the NJL limit, $\alpha\to0$, where the gauge interaction is eliminated. At the limit the running couplings (\ref{running:y}) and (\ref{running:lambda}) reduce to
\begin{align}
        &y_\Lambda^2(t) = \frac{16\pi^2}{2a}\frac{\alpha}{\alpha_c}\left[1 - \left(\frac{\mu^2}{\Lambda^2}\right)^{1-w}\right]^{-1} = -\frac{16\pi^2}{2a}\frac{2}{\ln(\mu^2/\Lambda^2)},
        \label{NJL:y}\\
        &\frac{\lambda_\Lambda(t)}{y_\Lambda^4(t)} = \frac{2a}{16\pi^2}\frac{\alpha_c}{\alpha}\left[1 - \left(\frac{\mu^2}{\Lambda^2}\right)^{2-2w}\right] = -\frac{2a}{16\pi^2}\ln(\mu^2/\Lambda^2).
        \label{NJL:lamda}
\end{align}
From Eqs.~(\ref{running:mass}) and (\ref{NJL:y}) we obtain
\begin{align}
        &m^2(t) = \frac{2a}{16\pi^2}\left(\frac{\Lambda^2}{\mu^2}\right)^w y^2_\Lambda (t) \mu^2 \left(\frac{1}{g_4(\Lambda)} - \frac{1}{w}\right)
         \xrightarrow{\Lambda\to\infty} \mu^2 \left(\frac{1}{g_{4R}(\mu)} - \frac{1}{g_{4R}^*}\right) ,
\end{align}
where we renormalize $g_4(\Lambda)$ suitably to obtain a finite expression \cite{Harada:1994wy}.

Following the similar procedure in Sec.~\ref{gNJL} we calculate the RG invariant effective potential at the limit $\Lambda\to\infty$
and obtain
\begin{align}
        V(\sigma) = \frac{1}{2}\left(\frac{1}{g_{4R}} - \frac{1}{g_{4R}^*}\right)\mu^2\sigma^2 + \frac{R}{12}\sigma^2 .
        \label{vj:NJL}
\end{align}
Thus, the gravitational effective action with the weak curvature $R\ll \sigma^2$ becomes
\begin{align}
        S^{(NJL)} = \int d^4x\sqrt{-g}
        \left[- \frac{1}{2}R + \frac{1}{2}g^{\mu\nu}\pd_\mu\sigma\pd_\nu\sigma
        - V(\sigma) \right].
\end{align}
By the Weyl transformation (\ref{Weyl}), one can eliminate the interaction term between the curvature and the auxiliary field $\sigma$.
In the Einstein frame the action is simplified as
\begin{align}
        S^{(NJL)}_{E} = \int d^4x\sqrt{-\tilde{g}}
        \left[ - \frac{1}{2}\tilde{R} + \frac{1}{2}\tilde{g}^{\mu\nu}\pd_\mu\varphi\pd_\nu\varphi
        - V_E(\sigma)\right] ,
\end{align}
with the effective potential
\begin{align}
        V_E(\sigma) =
        \frac{1}{2}\left[\frac{1}{g_{4R}} - \frac{1}{g_{4R}^*}\right]\frac{\mu^2\sigma^2}{(1 + \sigma^2/6)^2} .
        \label{ve:NJL}
\end{align}
The denominator $(1 + \sigma^2/6)^2$ comes from the NJL limit of $\Omega^2$ in Eq.~(\ref{omega}). It suppresses the effective potential for a large $\sigma$. The effective potential (\ref{ve:NJL}) has a maximum at $\sigma=\sqrt{6}$.

\begin{figure}[htbp]
\begin{minipage}{0.48\hsize}
        \centering
        \includegraphics[width=0.96\linewidth]{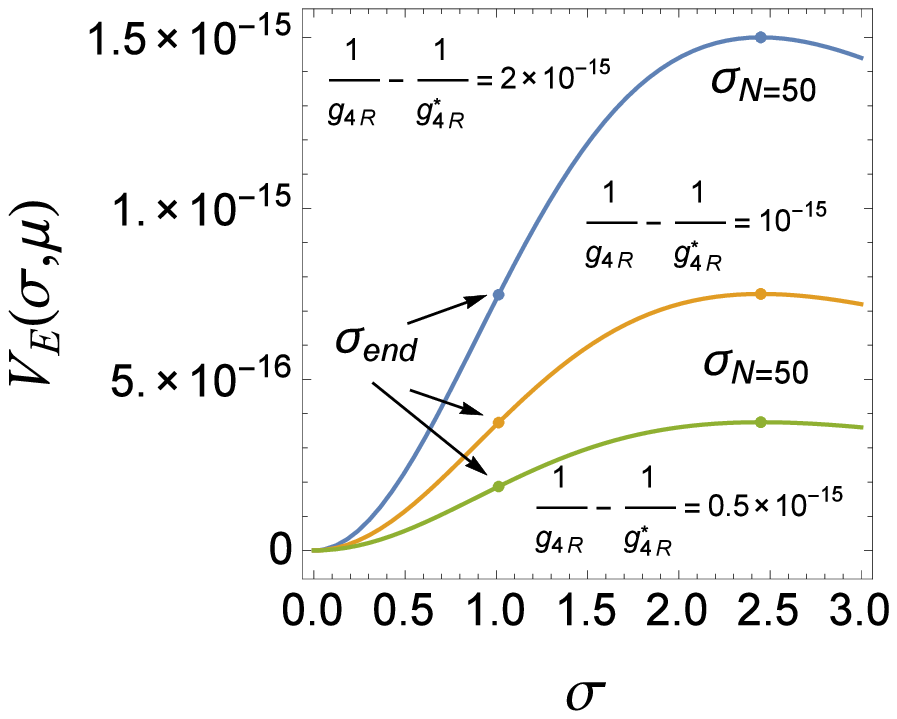}
        \vglue -4mm
        \caption{Behavior of the potential $V_E(\sigma)$ as a function of $\sigma$ for $\mu=1$ and $1/g_{4R}-1/g_{4R}^* = 0.5\times 10^{-15}, 1\times 10^{-15}, 2\times 10^{-15}$. It should be noticed that we do not adopt any model parameters whose effective potential has a local maximum except for this appendix.}
        \label{fig:potentialNJL}
\end{minipage}
\hspace{0.04\hsize}
\begin{minipage}{0.48\hsize}
        \centering
        \includegraphics[width=0.8\linewidth]{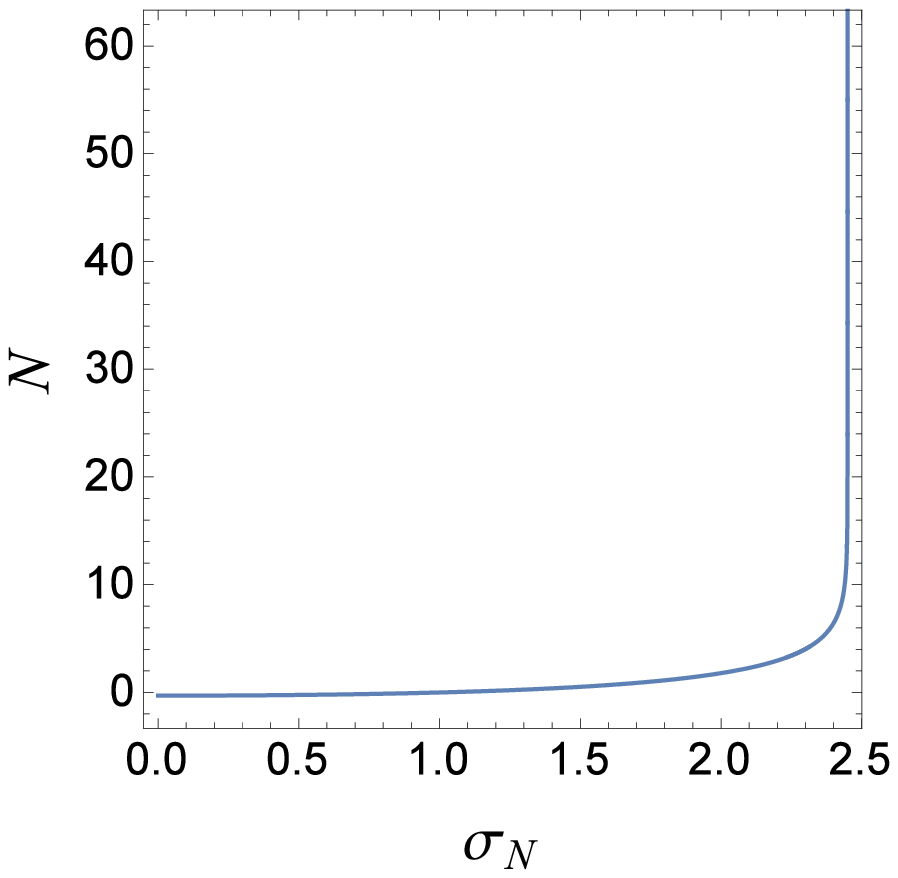}
        \vglue -4mm
        \caption{Behavior of the e-folding number $N$ as a function of $\sigma_N$.}
        \label{fig:efNJL}
\end{minipage}
\end{figure}

In Fig.~\ref{fig:potentialNJL}  the typical behavior of the effective potential (\ref{ve:NJL}) is shown.  It is observed that the horizon crossing $\sigma=\sigma_N$ should take place near the maximum of the potential to obtain large enough e-folding  $N=50\sim 60$. The e-folding  is plotted as a function of $\sigma_N$ in Fig.~\ref{fig:efNJL}. It blows up at $\sigma=\sqrt{6}$. We noted that the $g_{4R}$ dependence gives only a negligible contribution to the e-folding.  The result introduces a fine- tuning problem why $\sigma_N\sim \sqrt{6}$. To avoid this fine-tuning problem one should modify the potential. Since the potential can be easily modified by including terms which are neglected in our approximation, the fine-tuning
problem seems to be not so serious.



\begin{thebibliography}{9}
\addcontentsline{toc}{section}{References}
\bibitem{Nambu:1961}
  Y.~Nambu and G.~Jona-Lasinio,
  Phys.\ Rev.\  {\bf 122} (1961) 345;
  Phys.\ Rev.\  {\bf 124} (1961) 246.

\bibitem{Miransky:1993}
V.~A.~Miransky,
{\it Dynamical Symmetry Breaking in Quantum Field Theories},
World Scientific (1993); \\
M.~Harada and K.~Yamawaki,
  Phys.\ Rept.\  {\bf 381} (2003) 1
  [hep-ph/0302103].

\bibitem{Kondo:1991}
K.-I.~Kondo, S.~Shuto and K.~Yamawaki,
Mod.\ Phys.\ Lett.\ {\bf A6} (1991) 3385;\\
K.-I.~Kondo, M.~Tanabashi and K.~Yamawaki,
Progr.\ Theor.\ Phys.\ {\bf 89} (1993) 1249;
Mod.\ Phys.\ Lett.\ {\bf A8} (1993) 2859.

\bibitem{Leung:1985}
  C.~N.~Leung, S.~T.~Love and W.~A.~Bardeen,
  Nucl.\ Phys.\ B {\bf 273} (1986) 649;
  Nucl.\ Phys.\ B {\bf 323} (1989) 493.

\bibitem{Hill:2002ap}
  C.~T.~Hill and E.~H.~Simmons,
  Phys.\ Rept.\  {\bf 381} (2003) 235
  [Phys.\ Rept.\  {\bf 390} (2004) 553]
  [hep-ph/0203079].

\bibitem{Ade:2015tva}
  P.~A.~R.~Ade {\it et al.} [BICEP2 and Planck Collaborations],
  Phys.\ Rev.\ Lett.\  {\bf 114} (2015) 101301
  [arXiv:1502.00612 [astro-ph.CO]].

\bibitem{Ade:2015lrj}
  P.~A.~R.~Ade {\it et al.} [Planck Collaboration],
  arXiv:1502.01589 [astro-ph.CO].
  arXiv:1502.02114 [astro-ph.CO].

\bibitem{Inagaki:1993kz}
  T.~Inagaki, T.~Muta and S.~D.~Odintsov,
  Mod.\ Phys.\ Lett.\ A {\bf 8} (1993) 2117
  [hep-th/9306023];
Prog.\ Theor.\ Phys.\ Suppl.\  {\bf 127} (1997) 93
[hep-th/9711084].

\bibitem{Hill:1991jc}
  C.~T.~Hill and D.~S.~Salopek,
  Annals Phys.\  {\bf 213} (1992) 21.

\bibitem{Muta:1991mw}
  T.~Muta and S.~D.~Odintsov,
  Mod.\ Phys.\ Lett.\ A {\bf 6} (1991) 3641.

\bibitem{Geyer:1996wg}
B.~Geyer and S.~D.~Odintsov,
Phys.\ Lett.\ B {\bf 376} (1996) 260
[hep-th/9603172];
Phys.\ Rev.\ D {\bf 53} (1996) 7321
[hep-th/9602110].

\bibitem{Inagaki:2011jk}
  T.~Inagaki, Y.~Rybalov and X.~Meng,
  Eur.\ Phys.\ J.\ C {\bf 71} (2011) 1656
  [arXiv:1102.5036 [hep-ph]].

\bibitem{Iso:2014gka}
  S.~Iso, K.~Kohri and K.~Shimada,
  Phys.\ Rev.\ D {\bf 91}  (2015) 044006
  [arXiv:1408.2339 [hep-ph]].

\bibitem{eli}
 E.~Elizalde and S.~D.~Odintsov,
  Phys.\ Lett.\ B {\bf 303} (1993) 240
  [hep-th/9302074];
  Phys.\ Lett.\ B {\bf 321} (1994) 199
  [hep-th/9311087];Z.\ Phys.\ C {\bf 64} (1994) 699
  [hep-th/9401057].

\bibitem{rgi}
 A.~De Simone, M.~P.~Hertzberg and F.~Wilczek,
  Phys.\ Lett.\ B {\bf 678} (2009) 1
  [arXiv:0812.4946 [hep-ph]];\\
  S.~Mukaigawa, T.~Muta and S.~D.~Odintsov,
  Int.\ J.\ Mod.\ Phys.\ A {\bf 13} (1998) 2739
  [hep-ph/9709299];\\
 H.~M.~Lee,
  Phys.\ Lett.\ B {\bf 722} (2013) 198
  [arXiv:1301.1787 [hep-ph]]; \\
 N.~Okada and Q.~Shafi, arXiv:1311.0921 [hep-ph];
G.~Barenboim, E.~J.~Chun and H.~M.~Lee,
  Phys.\ Lett.\ B {\bf 730} (2014) 81;\\
  Y.~Hamada, H.~Kawai, K.~y.~Oda and S.~C.~Park,
  Phys.\ Rev.\ Lett.\  {\bf 112} (2014) 241301
  [arXiv:1403.5043 [hep-ph]]; \\
  R.~P.~Woodard,
  Phys.\ Rev.\ Lett.\  {\bf 101} (2008) 081301
  [arXiv:0805.3089 [gr-qc]];\\
 I.~Oda and T.~Tomoyose,
  Adv.\ Stud.\ Theor.\ Phys.\  {\bf 8} (2014) 551
  [arXiv:1404.1538 [hep-ph]];\\
J.~Ren, Z.~Z.~Xianyu and H.~J.~He,
  JCAP {\bf 1406} (2014) 032
  [arXiv:1404.4627 [gr-qc]];\\
  Y.~Hamada, H.~Kawai and K.~y.~Oda,
  JHEP {\bf 1407} (2014) 026;\\
  T.~Inagaki, R.~Nakanishi and S.~D.~Odintsov,
  Astrophys.\ Space Sci.\  {\bf 354} (2014) 2108
  [arXiv:1408.1270 [gr-qc]];
  Phys.\ Lett.\ B {\bf 745} (2015) 105
  [arXiv:1502.06301 [hep-ph]];\\
 E.~Elizalde, S.~D.~Odintsov, E.~O.~Pozdeeva and S.~Y.~Vernov,
  Phys.\ Rev.\ D {\bf 90} (2014) 084001
  [arXiv:1408.1285 [hep-th]]; \\
  Y.~Hamada, H.~Kawai, K.~y.~Oda and S.~C.~Park,
  Phys.\ Rev.\ D {\bf 91} (2015) 053008
  [arXiv:1408.4864 [hep-ph]]; \\
  H.~J.~He and Z.~Z.~Xianyu,
  JCAP {\bf 1410} (2014) 019
  [arXiv:1405.7331 [hep-ph]]; \\
  M.~Herranen, T.~Markkanen, S.~Nurmi and A.~Rajantie,
  Phys.\ Rev.\ Lett.\  {\bf 113} (2014) 211102
  [arXiv:1407.3141 [hep-ph]]; \\
  M.~Herranen, A.~Osland and A.~Tranberg,
  arXiv:1503.07661 [hep-ph].

\bibitem{Harada:1994wy}
M.~Harada, Y.~Kikukawa, T.~Kugo and H.~Nakano,
Prog.\ Theor.\ Phys.\  {\bf 92} (1994) 1161
[hep-ph/9407398].

\bibitem{Bardeen:1989ds}
  W.~A.~Bardeen, C.~T.~Hill and M.~Lindner,
  Phys.\ Rev.\ D {\bf 41} (1990) 1647.

\bibitem{Kaiser:1994}
  D.~I.~Kaiser,
  Phys.\ Lett.\ B {\bf 340} (1994) 23
  [astro-ph/9405029];
  Phys.\ Rev.\ D {\bf 52} (1995) 4295
  [astro-ph/9408044].

\bibitem{Linde:2005ht}
  A.~D.~Linde,
  Contemp.\ Concepts Phys.\  {\bf 5} (1990) 1
  [hep-th/0503203];\\
 D.~S.~Gorbunov and V.~A.~Rubakov,
 ``Introduction to the theory of the early universe: Cosmological perturbations and inflationary theory,''
  Hackensack, USA: World Scientific (2011) 489 p
  
\bibitem{Kohri:2013mxa}
  K.~Kohri, Y.~Oyama, T.~Sekiguchi and T.~Takahashi,
  JCAP {\bf 1310} (2013) 065
  [arXiv:1303.1688 [astro-ph.CO]].

\bibitem{buc}
I.~L.~Buchbinder, S.~D.~Odintsov, I.~L.~Shapiro, "{\it Effective Action in Quantum Gravity}," Bristol, UK: IOP (1992).

\bibitem{Holthausen:2013ota}
  M.~Holthausen, J.~Kubo, K.~S.~Lim and M.~Lindner,
  JHEP {\bf 1312} (2013) 076
  [arXiv:1310.4423 [hep-ph]].
 
  \bibitem{Bamba:2014}
  K.~Bamba, R.~Myrzakulov, S.~D.~Odintsov and L.~Sebastiani,
  Phys.\ Rev.\ D {\bf 90} (2014) 043505
  [arXiv:1403.6649 [hep-th]].
  

\end{thebibliography}

\end{document}